\documentclass{statsoc}
\usepackage{amsmath, amssymb}
\usepackage{graphicx}
\usepackage{cite, natbib}

\newcommand{\indicator}{\ensuremath{\mathbf{1}}}
\newcommand{\dif}[1][\:]{\ensuremath{#1 \mathrm{d}}}
\newcommand{\E}[2][]{\ensuremath{\mathbb{E} #1[\, #2 \, #1]}}
\newcommand{\condE}[3][]{\ensuremath{\mathbb{E} #1[\, #2 \, #1|\, #3 \, #1]}}
\newcommand{\Var}[2][]{\ensuremath{\mathrm{Var} #1( #2 #1)}}
\newcommand{\prodint}{\ensuremath{\text{{\huge $\pi$}}}}

\newcommand{\Sus}{\ensuremath{\mathcal{S}}}
\newcommand{\Inf}{\ensuremath{\mathcal{I}}}

\renewcommand{\th}{\ensuremath{^\text{th}}}

\title{Nonparametric survival analysis of epidemic data}
\author[Kenah]{Eben Kenah}
\address{Department of Biostatistics, University of Washington, Seattle, USA}
\email{eek4@u.washington.edu}

\begin{document}

\begin{abstract}
This paper develops nonparametric methods for the survival analysis of epidemic data based on contact intervals.  The contact interval from person $i$ to person $j$ is the time between the onset of infectiousness in $i$ and infectious contact from $i$ to $j$, where we define infectious contact as a contact sufficient to infect a susceptible individual.  We show that the Nelson-Aalen estimator produces an unbiased estimate of the contact interval cumulative hazard function when who-infects-whom is observed.  When who-infects-whom is not observed, we average the Nelson-Aalen estimates from all transmission networks consistent with the observed data using an EM algorithm.  This converges to a nonparametric MLE of the contact interval cumulative hazard function that we call the \textit{marginal Nelson-Aalen estimate}.  We study the behavior of these methods in simulations and use them to analyze household surveillance data from the 2009 influenza A(H1N1) pandemic.  In an appendix, we show that these methods extend chain-binomial models to continuous time.
\end{abstract}

\section{Introduction}
\label{sec:intro}
Infectious diseases remain one of the greatest threats to human health and commerce, and the analysis of epidemic data is one of the most important applications of statistics in public health.  If person $i$ infects person $j$ with a given disease, the generation interval is the time between the infection of $i$ and the infection of $j$.  The serial interval, which is often used as a proxy for the generation interval, is the time between the onset of symptoms in $i$ and the onset of symptoms in $j$.  Data from several recent and historical epidemics have been analyzed using methods based on generation or serial interval distributions, including the 1918 influenza \citep{Mills}, severe acute respiratory syndrome (SARS) \citep{LipsitchSARS, WallingaTeunis}, pandemic influenza A(H1N1) \citep{FraserH1N1, McBrydeH1N1, YangH1N1}, and avian influenza \citep{Ferguson1, Ferguson2}.  

Though the generation and serial interval distributions are often considered invariant features of an infectious disease \citep{Fine}, these distributions can change systematically over the course of an epidemic \citep{Svensson, Kenah3}.  When a susceptible person is exposed to multiple infectious people, his or her infector must be the first person to make infectious contact.  Thus, the mean generation and serial intervals contract as the prevalence of infection increases.  When transmission is rapid within groups of close contacts such as households or schools, this can occur even when the global prevalence of infection remains low \citep{Kenah3}.  

Statistical methods for infectious disease data that use generation or serial interval distributions are based on the Lotka-Euler equation \citep{WallingaLipsitch, RobertsHeesterbeek} or a branching-process approximation to the early spread of disease \citep{WallingaTeunis, WhitePagano}.  In both of these approaches, epidemic spread is modeled as a process that creates a population rather than a process of percolation through an existing population.  Thus, these methods fail to account for the effects of exposure to multiple infectors and ignore information contributed by uninfected person-time.  Since they assume generation intervals are independent and identically distributed, they implicitly assume a constant latent period (infection to onset of infectiousness) and a constant infectious period.  When serial intervals are used as a proxy for generation intervals, they implicitly assume that the incubation period (infection to onset of symptoms) is also constant.

\citet{Kenah4} outlined an alternative analysis of epidemic data based on \textit{contact intervals}.  Informally, the contact interval from an infectious person $i$ to a susceptible person $j$ is the time between the onset of infectiousness in $i$ and the first infectious contact from $i$ to $j$, where we define infectious contact to be a contact sufficient to infect a susceptible individual.  The contact interval is similar to the generation interval, but it is defined for all possible infector-infectee pairs, not just those in which infection is actually transmitted, and it begins with the onset of infectiousness, not infection.  The contact interval from $i$ to $j$ will be right-censored if $j$ is infected by $k\neq i$ prior to infectious contact from $i$ or if $i$ recovers before making infectious contact with $j$.  This censoring is accounted for using methods from survival analysis.  Statistical methods based on contact intervals avoid many of the problems with methods based on generation and serial intervals.  They can incorporate a greater variety of transmission models, they use information contributed by uninfected person-time, and they allow clearer expression of analytical assumptions. 

In epidemic modeling, it is common to specify infectious contact within an underlying contact process between individuals in the population.  While a person $i$ is infectious, he or she has some probability of transmitting infection to a susceptible person $j$ each time a contact from $i$ to $j$ is made~\citep[e.g.,][]{Rhodes}.  Models based on contact intervals define an infectious contact as a contact sufficient to infect a susceptible individual, ignore all contacts made by person $i$ outside his or her infectious period, and ignore all infectious contacts from $i$ to $j$ after the first.  This relaxes the assumption that the contact process is unaffected by infection and limits our attention to those contacts relevant to disease transmission.  Any model specified in terms of an underlying contact process can be specified equivalently in terms of contact intervals. 

This paper outlines the nonparametric survival analysis of epidemic data based on contact intervals.  The rest of Section~\ref{sec:intro} defines the general stochastic Susceptible-Exposed-Infectious-Removed (SEIR) model, describes our observed data, and reviews the parametric survival analysis of epidemic data.  Section~\ref{sec:methods} shows that the Nelson-Aalen estimator produces an unbiased estimate of the cumulative hazard function of the contact interval distribution when who-infected-whom is observed.  When who-infected-whom is not observed, the Nelson-Aalen estimates from all transmission networks consistent with the observed data are averaged to get a \emph{marginal Nelson-Aalen estimator}.  Since the transmission network probabilities are unknown, an expectation-maximization (EM) algorithm is used to iteratively reweight the possible transmission networks, producing a series of marginal Nelson-Aalen estimates that converges to a nonparametric maximum likelihood estimate (MLE) of the contact interval cumulative hazard function.  Section~\ref{sec:sims} explores the performance of these estimators in simulations.  Section~\ref{sec:data} uses them to analyze household surveillance data from the 2009 influenza A(H1N1) pandemic.  Section~\ref{sec:discussion} discusses the limitations, advantages, and future development of these methods.  Appendix~\ref{app:partialv} shows how our methods can be used when who-infected-whom is partially observed, and Appendix~\ref{app:chainbinom} shows how our methods generalize chain-binomial models to continuous time.

\subsection{Stochastic SEIR model and observed data}
Consider a closed population of $n$ individuals assigned indices $1\ldots n$.  Each individual is in one of four possible states: susceptible (S), exposed (E), infectious (I), or removed (R).  Person $i$ moves from S to E at his or her \textit{infection time} $t_i$, with $t_i = \infty$ if $i$ is never infected.  After infection $i$ has a \textit{latent period} of length $\varepsilon_i$, during which he or she is infected but not infectious.  At time $t_i +\varepsilon_i$, $i$ moves from E to I, beginning an \textit{infectious period} of length $\iota_i$.  At time $t_i + r_i$, where $r_i = \varepsilon_i + \iota_i$ is the \textit{recovery period}, $i$ moves from I to R.  Once in R, $i$ can no longer infect others or be infected.  The latent period is a nonnegative random variable, the infectious and recovery periods are strictly positive random variables, and the recovery period is finite with probability one.

An epidemic begins with one or more persons infected from outside the population, which we call \textit{imported infections}.  For simplicity, we assume that epidemics begin with one or more imported infections at time $0$ and there are no other imported infections.

After becoming infectious at time $t_i + \varepsilon_i$, person $i$ makes infectious contact with $j\neq i$ at time $t_{ij} = t_i + \varepsilon_i + \tau^*_{ij}$, where the \textit{infectious contact interval} $\tau^*_{ij}$ is a strictly positive random variable with $\tau^*_{ij} = \infty$ if infectious contact never occurs.  Since infectious contact must occur while $i$ is infectious or never, $\tau^*_{ij} \in (0, \iota_j]$ or $\tau^*_{ij} = \infty$.  We define infectious contact to be sufficient to cause infection in a susceptible person, so $t_j\leq t_{ij}$.

For each ordered pair $ij$, let $C_{ij} = 1$ if infectious contact from $i$ to $j$ is possible and $C_{ij} = 0$ otherwise.  We assume that the infectious contact interval $\tau^*_{ij}$ is generated in the following way: A \textit{contact interval} $\tau_{ij}$ is drawn from a distribution with hazard function $\lambda_{ij}(\tau)$.  If $\tau_{ij}\leq\iota_i$ and $C_{ij} = 1$, then $\tau^*_{ij} = \tau_{ij}$.  Otherwise, $\tau^*_{ij} = \infty$.  In this paper, we assume that all contact intervals have a continuous distribution and, for a fixed $i$ or a fixed $j$, the contact intervals $\tau_{ij}$, $j\neq i$, are independent.  

The parametric methods in \citet{Kenah4} are derived using counting processes and martingales, which are described in \citet{KalbfleischPrentice} and \citet{Aalen}.  Let $\Sus_i(t) = \mathbf{1}_{t\leq t_i}$ and $\Inf_i(t) = \mathbf{1}_{t\in (t_i + \varepsilon_i, t_i + r_i]}$ be the susceptibility and infectiousness processes for person $i$, where $\mathbf{1}_X = 1$ if $X$ is true and zero otherwise.  These processes are left-continuous and infectious contact from $i$ to $j$ is possible at time $t$ only if $C_{ij}\Inf_i(t)\Sus_j(t) = 1$.

Following \citet{WallingaTeunis}, let $v_j$ denote the index of the person who infected person $j$, with $v_j = 0$ for imported infections and $v_j = \infty$ for persons not infected at or before time $T$.  The \textit{transmission network} is the directed network with an edge from $v_j$ to $j$ for each $j$ such that $t_j\leq T$.  It can be represented by a vector $\mathbf{v} = (v_1,\ldots, v_n)$.  Let $\mathcal{V}_j = \{i: C_{ij}I_i(t_j) = 1\}$ denote the set of persons who could have infected person $j$, which we call this the \textit{infectious set} of person $j$.  Let $\mathcal{V}$ denote the set of all possible $\mathbf{v}$ consistent with the observed data.

Our population has size $n$, and we observe the times of all $\text{S}\rightarrow\text{E}$ (infection), $\text{E}\rightarrow\text{I}$ (onset of infectiousness), and $\text{I}\rightarrow\text{R}$ (removal) transitions in the population between time $0$ and time $T$.  For all ordered pairs $ij$ such that $i$ is infected, we observe $C_{ij}$.  Let $m$ denote the total number of infections we observe.  Our goal is to estimate the contact interval distribution via its cumulative hazard function.

\subsection{Parametric survival analysis of epidemic data}
We begin by reviewing the parametric survival analysis of epidemic data \citep{Kenah4}, where we assume that $\lambda_{ij}(\tau)$ comes from a family of hazard functions indexed by a parameter vector $\theta$ with a unique $\theta_0$ such that $\lambda_{ij}(\tau) = \lambda(\tau;\theta_0)$ for all $ij$.  To guarantee that the maximum likelihood estimates are consistent and asymptotically normal, we assume that, for all $\tau$, $\lambda(\tau;\theta)$ and $\ln\lambda(\tau;\theta)$ have continuous third derivatives with respect to $\theta$ in an open neighborhood of $\theta_0$.

\subsubsection{Likelihood when who-infects-whom is observed} 
Let $\mathcal{N}_{ij}(t) = \mathbf{1}_{t\geq t_{ij}}$ count the first infectious contact from $i$ to $j$.  By definition, $\mathcal{N}_{ij}(t)$ is right-continuous with left-hand limits (cadlag).  We count only the first infectious contact because $j$ is infected at or before $t_{ij}$.  Since $\lambda_{ij}(\tau) = \lambda(\tau;\theta_0)$ and $\mathcal{N}_{ij}(0) = 0$, the process
\begin{equation}
    \mathcal{M}_{ij}(t;\theta) = \mathcal{N}_{ij}(t) - \int_0^t \lambda(u - t_i - \varepsilon_i;\theta)C_{ij}\Inf_i(u)\indicator_{u\leq t_{ij}}\dif u
\end{equation}	
is a zero-mean martingale when $\theta = \theta_0$.  Since $\Sus_j(u)\mathbf{1}_{u\leq T}$ is predictable, the stochastic integral
\begin{equation}
    M_{ij}(t;\theta) = \int_0^t \Sus_j(u)\mathbf{1}_{u\leq T}\dif\mathcal{M}_{ij}(u;\theta)
\end{equation}
is a zero-mean martingale when $\theta = \theta_0$.  The corresponding counting process is
\begin{equation}
    N_{ij}(t) = \int_0^t \Sus_j(u)\mathbf{1}_{u\leq T}\dif\mathcal{N}_{ij}(u),
\end{equation}
which counts infectious contacts from $i$ to $j$ while $j$ is susceptible and before time $T$.  The corresponding log likelihood is
\begin{equation}
    \ell_{ij}(\theta) = \int_0^T \ln\lambda(u - t_i - \varepsilon_i;\theta)\Sus_j(u)\,dN_{ij}(u) - \int_0^T \lambda(u - t_i - \varepsilon_i;\theta)C_{ij}\Inf_i(u)\Sus_j(u)\dif u,
\end{equation}
which has the score process 
\begin{equation}
    U_{ij}(t; \theta) = \int_0^t \frac{\partial}{\partial\theta}\ln \lambda(u - t_i - \varepsilon_i;\theta)\,dM_{ij}(u;\theta).
\end{equation}
Since it is the integral of a predictable process with respect to a zero-mean martingale, $U_{ij}(t;\theta_0)$ is a zero-mean martingale.  If $t_j\leq T$, this likelihood requires that we observe whether or not $i$ made infectious contact with $j$ at time $t_j$.

Now fix $j$.  If we observe infectious contacts in all ordered pairs $ij$ from time $0$ until time $T$, the log likelihood is
\begin{equation}
    \ell_{\cdot j}(\theta) = \sum_{i\neq j} \ell_{ij}(\theta),
    \label{lj}
\end{equation}
with the score process
\begin{equation}
    U_{\cdot j}(t;\theta) = \sum_{i\neq j} U_{ij}(t;\theta).
    \label{Uj}
\end{equation}
$U_{\cdot j}(t;\theta_0)$ is a zero-mean martingale because it is a sum of zero-mean martingales.  If $t_j\leq T$, the likelihood in equation \eqref{lj} requires that we observe which counting process $N_{ij}$ jumped at $t_j$, which is equivalent to observing which $i$ infected $j$.  The complete-data log likelihood when we observe who-infects-whom is $\ell(\theta) = \sum_{j = 1}^n \ell_{\cdot j}(\theta)$ with the score process $U(t;\theta) = \sum_{j = 1}^n U_{\cdot j}(t;\theta)$.  Clearly, $U(t;\theta_0)$ is a zero-mean martingale.

\subsubsection{Likelihood when who-infects-whom is not observed}
Now assume that we do not observe who infected person $j$.  The corresponding counting process is
\begin{equation}
    N_{\cdot j}(t) = \sum_{i\neq j} N_{ij}(t),
    \label{Nj}
\end{equation}
and the process $M_{\cdot j}(t;\theta) = \sum_{i\neq j} M_{ij}(t;\theta)$ is a zero-mean martinale when $\theta = \theta_0$.  Expanding $M_{\cdot j}(t;\theta)$, we get
\begin{equation}
    M_{\cdot j}(t;\theta) = N_{\cdot j}(t) - \int_0^t \lambda_{\cdot j}(u;\theta)\Sus_j(u)\mathbf{1}_{u\leq T}\dif u,
\end{equation}
where
\begin{equation}
    \lambda_{\cdot j}(t;\theta) = \sum_{i\neq j} \lambda(t - t_i - \varepsilon_i; \theta)C_{ij}\Inf_i(t),
    \label{lambdaj}
\end{equation}
is the total hazard of infectious contact with $j$ at time $t$ given $\theta$.  When person $j$ is observed from time $0$ to time $T$, the log likelihood is
\begin{equation}
    \widetilde{\ell}_{\cdot j}(\theta) = \int_0^T \ln\lambda_{\cdot j}(u;\theta)\,dN_{\cdot j}(u) - \int_0^T \lambda_{\cdot j}(u;\theta)\Sus_j(u)\,du
    \label{tildelj}
\end{equation}
with the score process
\begin{equation}
    \widetilde{U}_{\cdot j}(t;\theta) = \int_0^t \frac{\partial}{\partial\theta}\ln\lambda_{\cdot j}(u;\theta)\,dM_{\cdot j}(u;\theta).
    \label{tildeUj}
\end{equation}
$\widetilde{U}_{\cdot j}(t;\theta_0)$ is a zero-mean martingale because it is the integral of a predictable process with respect to $M_{\cdot j}(t;\theta_0)$.  The log likelihood for the complete observed data when we do not observe who-infected-whom is $\widetilde{\ell}(\theta) = \sum_{j=1}^n \widetilde{\ell}_{\cdot j}(\theta)$ with score process $\widetilde{U}(t;\theta) = \sum_{j=1}^n \widetilde{U}_{\cdot j}(t;\theta)$, which is a zero-mean martingale when $\theta = \theta_0$.

\section{Methods}
\label{sec:methods}
In this section, we extend the Nelson-Aalen estimator \citep{Altshuler, Nelson, Aalen1978} to derive a nonparametric estimators of the cumulative hazard function of the contact interval distribution.  We continue to assume that the contact interval distribution is continuous, but we drop the dependence of the hazard function on a parameter vector $\theta$, so it is simply $\lambda(\tau)$.  Let 
\begin{equation}
    \Lambda(\tau) = \int_0^\tau \lambda(u)\dif u
\end{equation}
denote the cumulative hazard function.  We call $\tau$ the \emph{infectiousness age}.  To derive parametric estimators of the contact interval distribution, we used counting processes and martingales defined in absolute time.  To derive nonparametric estimators, we use counting processes and martingales defined in infectiousness age.  Whereas absolute time processes are defined for all ordered pairs $ij$, infectiousness age processes are defined only for those $ij$ in which $t_i <\infty$.  In our notation, we use the argument $t$ for absolute time and $\tau$ for infectiousness age.  Martingales and counting processes in infectiousness age are given an asterisk superscript.  

We show that the Nelson-Aalen estimator generates an unbiased estimate of $\Lambda(\tau)$ when who-infects-whom is observed.  When who-infects-whom is not observed, we average the Nelson-Aalen estimates from all $\mathbf{v}\in\mathcal{V}$ to get a \emph{marginal Nelson-Aalen estimator}.  Since the probability of each $\mathbf{v}\in\mathcal{V}$ is unknown, we use an EM algorithm \citep{DempsterLairdRubin} to iteratively reweight each possible $\mathbf{v}$, obtaining a sequence of marginal Nelson-Aalen estimates that converges to a nonparametric MLE of $\Lambda(\tau)$.  Finally, we show how these estimates can be approximated in mass-action SEIR models using only data about infected persons.  In Appendix~\ref{app:partialv}, we show how these methods adapt to a situation where the transmission network is partially observed.

\subsection{Nonparametric estimation when who-infects-whom is observed}
First, we assume that we observe who infected whom.  For each person $i$ such that $t_i +\varepsilon_i\leq T$, let $\mathcal{N}^*_{ij}(\tau) = \indicator_{\tau\geq\tau_{ij}}$ count the first infectious contact from $i$ to $j$ at or before infectiousness age $\tau$ in person $i$, and let $\Inf^*_i(\tau) = \indicator_{\tau\in (0, \iota_i]}$ indicate whether $i$ remains infectious at infectiousness age $\tau$.  By definition, $\mathcal{N}^*_{ij}(\tau)$ is cadlag and $\Inf_i(\tau)$ is left-continuous.  Since $\mathcal{N}_{ij}(0) = 0$,
\begin{equation}
    \mathcal{M}^*_{ij}(\tau) = \mathcal{N}^*_{ij}(\tau) - \int_0^\tau \lambda(u)C_{ij}\Inf^*_i(u)\dif u
\end{equation}
is a zero-mean martingale.  Now let $\Sus_{ij}^*(\tau) = \Sus_j(t_i +\varepsilon_i + \tau)$ indicate the susceptibility of person $j$ at infectiousness age $\tau$ of person $i$ and let $\mathcal{T}_i = T - t_i - \varepsilon_i$ denote the time between the onset of infectiousness in $i$ and the end of observation.  Since $\Sus_{ij}^*(\tau)\indicator_{\tau\leq\mathcal{T}_i}$ is predictable, 
\begin{equation}
    M_{ij}^*(\tau) = \int_0^\tau \Sus_{ij}^*(u)\indicator_{u\leq\mathcal{T}_i}\dif\mathcal{M}^*_{ij}(u)
\end{equation}
is a zero-mean martingale corresponding to the counting process
\begin{equation}
    N_{ij}^*(\tau) = \int_0^\tau \Sus_{ij}^*(u)\indicator_{u\leq\mathcal{T}_i}\dif\mathcal{N}^*_{ij}(u),
\end{equation}
which counts infectious contacts from $i$ to $j$ prior to time $T$ while $j$ remains susceptible.

Now fix $j$.  Since it is a sum of zero-mean martingales, 
\begin{equation}
    M_{\cdot j}^*(\tau) = \sum_{i: t_i + \varepsilon_i\leq T} M_{ij}^*(\tau) = N^*_{\cdot j}(\tau) - \int_0^\tau \lambda(u)Y_{\cdot j}(u)\dif u,
\end{equation}
is a zero-mean martingale, where 
\begin{equation}
    N_{\cdot j}^*(\tau) = \sum_{i: t_i + \varepsilon_i\leq T} N_{ij}^*(\tau)\; =\;\indicator_{\tau\geq t_j - t_{v_j}}
    \label{eq:Nj}
\end{equation}
is a counting process that jumps at the infectiousness age at which $v_j$ (the infector of $j$) makes infectious contact with $j$ and 
\begin{equation}
    Y_{\cdot j}(\tau) = \sum_{i: t_i + \varepsilon_i\leq T} C_{ij}\Inf_i^*(\tau)\Sus_{ij}^*(\tau)\indicator_{\tau\leq\mathcal{T}_i}
    \label{eq:Yj}
\end{equation}
denotes the number of persons who could have infected $j$ at infectiousness age $\tau$.  Note that $Y_{\cdot j}(\tau)$ is left-continuous and decreasing in $\tau$.

Finally, consider the combined observations of all $j$.  Since it is a sum of zero-mean martingales,
\begin{equation}
    M^*(\tau) = \sum_{j=1}^n M_{\cdot j}^*(\tau) = N^*(\tau) - \int_0^\tau \lambda(u)Y(u)\dif u
\end{equation}
is a zero-mean martingale, where
\begin{equation}
    N(\tau) = \sum_{j=1}^n N^*_{\cdot j}(\tau)
\end{equation}
counts the number of observed infectious contacts with susceptible individuals occuring at infectiousness age $\leq\tau$ and
\begin{equation}
    Y(\tau) = \sum_{j=1}^n Y_{\cdot j}(\tau)
    \label{eq:Y}
\end{equation}
denotes the number of contact intervals of length $\geq\tau$ that were observed.  Like each $Y_{\cdot j}(\tau)$, $Y(\tau)$ is left-continuous and decreasing in $\tau$.

We are now ready to show that the standard Nelson-Aalen estimator will produce an unbiased estimate of $\Lambda(\tau)$.  Let
\begin{equation}
    \hat{\Lambda}(\tau) = \int_0^\tau \frac{\mathbf{1}_{Y(u) > 0}}{Y(u)}\dif N^*(u)
    \label{eq:hatLambda}
\end{equation}
be the Nelson-Aalen estimate of $\Lambda(\tau)$ and let $\mathcal{T} = \max\{\tau:\;Y(\tau) > 0\}$.  Then
\begin{equation}
    \hat{\Lambda}(\tau) - \Lambda(\tau\wedge\mathcal{T}) = \int_0^\tau \frac{\mathbf{1}_{Y(u)>0}}{Y(u)}\dif M^*(u),
\end{equation}
is a zero-mean martingale, where $\tau\wedge\mathcal{T} = \min(\tau, \mathcal{T})$ .  Therefore, $\hat{\Lambda}(\tau)$ is an unbiased estimate of $\Lambda(\tau)$ for all $\tau\in[0, \mathcal{T}]$.  For these $\tau$, a variance estimate for $\hat{\Lambda}(\tau) - \Lambda(\tau)$ is given by its optional variation process
\begin{equation}
    \hat{\sigma}^2(\tau) = \int_0^\tau \frac{\mathbf{1}_{Y(u)>0}}{Y(u)^2}\dif N(u).
    \label{eq:hatsigma}
\end{equation}
Using the martingale central limit theorem and a log transformation, we get the approximate pointwise $1 - \alpha$ confidence limits
\begin{equation}
    \hat{\Lambda}(\tau)\exp\Big(\pm \frac{\hat{\sigma}(\tau)}{\hat{\Lambda}(\tau)}\Phi^{-1}\big(1 - \frac{\alpha}{2}\big)\Big),
    \label{eq:hatLambdaCI}
\end{equation} 
where $\Phi$ is the cumulative distribution function of the standard normal distribution.  Note that the point estimate $\hat{\Lambda}(\tau)$ in equation~\eqref{eq:hatLambda} is valid for an arbitrary contact interval distribution but the variance estimate $\hat{\sigma}^2(\tau)$ in equation~\eqref{eq:hatsigma} and the approximate confidence interval in equation~\eqref{eq:hatLambdaCI} assume a continuous contact interval distribution.

\subsection{Nonparametric estimation when who-infects-whom is not observed}
\label{sec:EMalgorithm}
Now assume that we do not observe who-infected-whom.  We can no longer calculate the Nelson-Aalen estimate in equation~\eqref{eq:hatLambda} because we do not know which of the contact intervals are censored.  However, 
\begin{equation}
    \Lambda(\tau) = \E[\big]{\hat{\Lambda}(\tau)} = \E[\Big]{\condE[\big]{\hat{\Lambda}(\tau)}{\text{observed data}}}
\end{equation}
by the law of iterated expectation, so we can still obtain an unbiased estimate of $\Lambda(\tau)$.  Let $\hat{\Lambda}_\mathbf{v}(\tau)$ denote the value of $\hat{\Lambda}(\tau)$ that we would have calculated had we observed the transmission network $\mathbf{v}$.  Then
\begin{equation}
    \widetilde{\Lambda}(\tau) = \sum_{\mathbf{v}\in\mathcal{V}} \hat{\Lambda}_\mathbf{v}(\tau)\Pr(\mathbf{v}|\text{observed data})
    \label{eq:tildeLambda1}
\end{equation}
is an unbiased estimate of $\Lambda(\tau)$ for each $\tau\in[0, \mathcal{T}]$.  For each $\mathbf{v}\in\mathcal{V}$, $\Pr(\mathbf{v}|\text{observed data})$ can be calculated if we know $\lambda(\tau)$ \citep{Kenah3}.  For each person $j$ with $t_j <\infty$, the probability that $j$ was infected by person $i\in\mathcal{V}_j$ is
\begin{equation}
    p_{ij} = \frac{\lambda(t_j - t_i - \varepsilon_i)}{\sum_{k\in\mathcal{V}_j}\lambda(t_j - t_k - \varepsilon_k)},
    \label{eq:pij}
\end{equation}
and the probability of a given transmission network $\mathbf{v} = (v_1,\ldots, v_n)$ is
\begin{equation}
    \Pr(\mathbf{v}|\text{observed data}) = \prod_{j:\, t_j <\infty} p_{v_j j},
    \label{eq:Prv}
\end{equation}
so the infector $v_j\in\mathcal{V}_j$ of each infected $j$ can be chosen independently of all $v_i$, $i\neq j$.  Note that equations \eqref{eq:pij} and \eqref{eq:Prv} assume a continuous contact interval distribution, which ensures that simultaneous infectious contacts have probability zero.  

With a known or estimated $\lambda(\tau)$, it is easy (but unnecessary) to calculate a marginal Nelson-Aalen estimate.  We outline this calculation and then use it as the basis of an EM algorithm that converges to a nonparametric MLE of $\Lambda(\tau)$.  First, let
\begin{equation}
    \widetilde{N}^*_{\cdot j}(\tau|\lambda) = \condE[\big]{N_{\cdot j}^*(\tau)}{\text{observed data}, \lambda},
    \label{eq:tildeNj}
\end{equation}
which is a cadlag step function with a jump of size $p_{ij}$ at $\tau = t_j - t_i - \varepsilon_i$ for each $i\in\mathcal{V}_j$.  Now let
\begin{equation}
    \widetilde{N}^*(\tau|\lambda) = \sum_{j=1}^n \widetilde{N}^*_{\cdot j}(\tau|\lambda).
\end{equation}
By equations \eqref{eq:tildeLambda1} through \eqref{eq:Prv}, the marginal Nelson-Aalen estimate given $\lambda(\tau)$ is
\begin{equation}
    \widetilde{\Lambda}(\tau|\lambda) = \int_0^\tau \frac{\indicator_{Y(u)>0}}{Y(u)}\dif\widetilde{N}(u|\lambda).
    \label{eq:tildeLambda}
\end{equation}
If $\lambda(\tau)$ is the true contact interval hazard function, this is an unbiased estimate of $\Lambda(\tau)$ for each $\tau\in [0, \mathcal{T}]$.  The whole point of calculating $\widetilde{\Lambda}(\tau)$ is that $\lambda(\tau)$ is unknown, but equation~\eqref{eq:tildeLambda} can be used in the following EM algorithm:
\begin{enumerate}
    \item Begin with an initial $\lambda^{(0)}(\tau)$.  This could be a parametric estimate or a constant.  Use this to calculate an initial marginal Nelson-Aalen estimate $\widetilde{\Lambda}(\tau|\lambda^{(0)})$.
    \item \label{step2} Use $\widetilde{\Lambda}(\tau|\lambda^{(k)})$ to calculate a new hazard function estimate $\lambda^{(k+1)}(\tau)$.
    \item \label{step3} Update the probabilities in equation \eqref{eq:pij} using $\lambda^{(k+1)}(\tau)$.
    \item \label{step4} Use the updated probabilities to calculate $\widetilde{\Lambda}(\tau|\lambda^{(k+1)})$ using equation \eqref{eq:tildeLambda}.
    \item Repeat the smoothing step~(\ref{step2}), the E-step~(\ref{step3}) and the M-step~(\ref{step4}) until the sequence $\big(\widetilde{\Lambda}(\tau|\lambda^{(k)})\big)$ converges.
\end{enumerate}  
The limit of this sequence is the marginal Nelson-Aalen estimate, which we denote $\widetilde{\Lambda}(\tau)$.  To show that this is truly an EM algorithm, we must prove that $\widetilde{\Lambda}(\tau|\lambda^{(k)})$ maximizes the expected log likelihood given $\lambda^{(k)}(\tau)$.  To do this, we adapt the argument of \citet{KaplanMeier} that their product-limit estimate of the survival function is a nonparametric MLE.

Let $\tau_1,\ldots, \tau_M$ denote the distinct infectiousness ages at which infectious contacts might have been made, and let $S(\tau)$ denote an arbitrary survival function for the contact interval distribution.  The conditional probability of failure (i.e., infectious contact) at infectiousness age $\tau_j$ given survival (i.e., no infectious contact) until $\tau_j$ is
\begin{equation}
    p_j = \frac{S(\tau_j^-) - S(\tau_j)}{S(\tau_j^-)},
\end{equation}
where $S(\tau_j^-) = \lim_{\tau\uparrow\tau_j} S(\tau)$ exists because $S(\tau)$ is cadlag.  When who-infects-whom is not observed, the expected log likelihood given $\lambda^{(k)}$ is
\begin{equation}
    G(S|\lambda^{(k)}) = \sum_{j=1}^M \Big(\ln p_j\dif[\,] N^*(\tau_j|\lambda^{(k)}) + \ln(1 - p_j)\big[Y(\tau_j) - \dif[] N^*(\tau_j|\lambda^{(k)})\big]\Big).
\end{equation}
The $j^\text{th}$ term of $G(S)$ is maximized by
\begin{equation}
    \widetilde{p}_j = \frac{\dif[]N^*(\tau_j|\lambda^{(k)})}{Y(\tau_j)},
\end{equation}
so the survival function that maximizes $G(S)$ is
\begin{align}
    \widetilde{S}(\tau) &= \prod_{j: \tau_j\leq\tau} (1 - \widetilde{p}_j)\\
    &= \prodint_0^\tau \Big(1 - \frac{\indicator_{Y(u)>0}}{Y(u)}\dif N^*(u|\lambda^{(k)})\Big), 
    \label{eq:tildeS}
\end{align}
where $\prodint$ represents the product integral described in \citet{KalbfleischPrentice} and \citet{Aalen}.  The Nelson-Aalen estimate corresponding to the Kaplan-Meier estimate in equation~\eqref{eq:tildeS} is 
\begin{equation}
    \widetilde{\Lambda}(\tau|\lambda^{(k)}) = \int_0^\tau \frac{\indicator_{Y(u)>0}}{Y(u)}\dif N^*(\tau|\lambda^{(k)}).
\end{equation}
Therefore, each iteration of the proposed EM algorithm maximizes the expected log likelihood, so the marginal Nelson-Aalen estimate is a nonparametric MLE of $\Lambda(\tau)$.  The corresponding marginal Kaplan-Meier estimator is a nonparametric MLE of $S(\tau)$.

The variance $\widetilde{\sigma}^2(\tau)$ of $\widetilde{\Lambda}(\tau)$ can be estimated using the conditional variance formula.  Conditioning on the transmission network $\mathbf{v}$, we get
\begin{equation}
    \widetilde{\sigma}^2(\tau) = \E[\big]{\hat{\sigma}^2_\mathbf{v}(\tau)} + \Var[\big]{\hat{\Lambda}_\mathbf{v}(\tau)},
    \label{eq:condvar}
\end{equation}
where $\hat{\Lambda}_\mathbf{v}(\tau)$ is the Nelson-Aalen estimate from equation~\eqref{eq:hatLambda} and $\hat{\sigma}^2_\mathbf{v}(\tau)$ is the variance estimate from equation~\eqref{eq:hatsigma} that we would have calculated had we observed the transmission network $\mathbf{v}$.  Let $\widetilde{\lambda}(\tau)$ denote the hazard function estimate corresponding to $\widetilde{\Lambda}(\tau)$, and let $\widetilde{N}^*(\tau) = \widetilde{N}^*\big(\tau\big|\widetilde{\lambda}\big)$.  The first term of equation~\eqref{eq:condvar} reduces to
\begin{equation}
    \E[\big]{\hat{\sigma}^2_\mathbf{v}(\tau)} = \int_0^\tau \frac{\indicator_{Y(u) > 0}}{Y(u)^2}\dif\widetilde{N}^*(u), 
\end{equation}
and the second term reduces to
\begin{equation}
    \Var[\big]{\hat{\Lambda}_\mathbf{v}(\tau)} = \int_0^\tau \frac{\indicator_{Y(u)>0}}{Y(u)^2}\dif\widetilde{N}^*(u)\; - \sum_{j:\,t_j <\infty} \Big(\int_0^\tau \frac{\indicator_{Y(u)>0}}{Y(u)}\dif\widetilde{N}^*_{\cdot j}(u)\Big)^2,
\end{equation}
where $\widetilde{N}^*_{\cdot j}(\tau) = \widetilde{N}^*_{\cdot j}\big(\tau\big|\widetilde{\lambda}\big)$.  Therefore,
\begin{equation}
    \widetilde{\sigma}^2(\tau) = 2\int_0^\tau \frac{\indicator_{Y(u)>0}}{Y(u)^2}\dif\widetilde{N}^*(u)\; - \sum_{j: t_j <\infty} \Big(\int_0^\tau \frac{\indicator_{Y(u)>0}}{Y(u)}\dif\widetilde{N}^*_{\cdot j}(u)\Big)^2
    \label{eq:tildesigma}
\end{equation}
Using the martingale central limit theorem and a log transformation, we get the approximate pointwise $1 - \alpha$ confidence limits
\begin{equation}
    \widetilde{\Lambda}(\tau)\exp\Big(\pm\frac{\widetilde{\sigma}(\tau)}{\widetilde{\Lambda}(\tau)}\Phi^{-1}\big(1 - \frac{\alpha}{2}\big)\Big).
    \label{eq:tildeLambdaCI}
\end{equation}

\subsection{Estimating the hazard function}
Obtaining $\lambda^{(k+1)}(\tau)$ from $\widetilde{\Lambda}(\tau|\lambda^{(k)})$ in step~(\ref{step2}) of the EM algorithm is nontrivial.  In general, the increments of $\widetilde{\Lambda}(\tau|\lambda^{(k)})$ cannot be used directly as $\lambda^{(k+1)}(\tau)$.  To see this, consider a person $j$ with $t_j\leq T$.  For each person $i\in\mathcal{V}_j$, let $p_{ij}^{(k)}$ denote the estimated probability that $i$ infected $j$ in the $k^\text{th}$ iteration of the EM algorithm, with $p_{ij}^{(0)}$ denoting the initial estimate.  In the first iteration, we get $p_{ij}^{(1)}\propto p_{ij}^{(0)}/Y(\tau_{ij})$.  After $k$ iterations,
\begin{equation}
    p_{ij}^{(k+1)} \propto \frac{p_{ij}^{(0)}}{Y(\tau_{ij})^{k+1}}.
\end{equation}
Since $Y(\tau)$ is decreasing in $\tau$, the EM algorithm would converge to a marginal Nelson-Aalen estimate where the $i\in\mathcal{V}_j$ with the longest $\tau_{ij}$ is assigned probability one of being $v_j$.  This problem arises because equation~\eqref{eq:pij} assumes a continuous contact interval distribution, which ensures that simultaneous infectious contacts have zero probability.  Since the contact interval distribution is continuous, the intervals between the increments of $\widetilde{\Lambda}(\tau|\lambda^{(k)})$ contain information about $\lambda(\tau)$ that is ignored if these increments are used directly as an estimate of $\lambda(\tau)$.  Thus, obtaining $\lambda^{(k+1)}(\tau)$ from $\Lambda(\tau|\lambda^{(k)})$ requires some form of smoothing.  

The algorithm depends on the smoothed estimate of $\lambda(\tau)$ only to calculate the probabilities $p_{ij}$ in equation~\eqref{eq:pij}.  In our experience, cubic smoothing splines and kernel smoothers produce nearly identical results, so the algorithm seems insensitive to reasonable choices among smoothers.

\subsection{Approximate Nelson-Aalen estimates for mass-action SEIR models}
\label{sec:massaction}
In a mass-action model, $C_{ij} = 1$ for all ordered pairs $ij$ but the hazard of infectious contact is inversely proportional to the population size $n$.  More specifically, if $\Lambda_{n}(\tau)$ is the cumulative hazard function of the contact interval distribution in a population of size $n$, then
\begin{equation}
    \Lambda_n(\tau) = \frac{\Lambda_*(\tau)}{n-1},
\end{equation}
where we call $\Lambda_*(\tau)$ the \textit{normalized cumulative hazard function} of the contact interval distribution.  With knowledge of $n$, the methods of the previous two sections could be used to estimate $\Lambda_n(\tau)$ and $\Lambda_*(\tau)$.  In \citet{Kenah4}, it was shown that the full parametric likelihoods for mass-action SEIR models can be approximated by likelihoods that depend only on data about infected individuals.  Here, we derive approximate nonparametric estimates of $\Lambda_*(\tau)$ that depend only on data about infected individuals.  For a fixed number $m$ of infections, these approximations become exact in the limit as $n\rightarrow\infty$.  Thus, they can be used to analyze data from the early stages of an epidemic, when $m\ll n$.

Let $Y_*(\tau) = \sum_i I_i^*(\tau)\mathbf{1}_{\tau\leq\mathcal{T}_i}$ denote the number of infected persons who remain infectious and under observation at infectiousness age $\tau$.  By equations \eqref{eq:Yj} and \eqref{eq:Y},
\begin{equation}
    Y(\tau) = \sum_{i:t_i +\varepsilon_i\leq T} \Big(I_i^*(\tau)\mathbf{1}_{\tau\leq\mathcal{T}_i}\sum_{j:j\neq i} S_{ij}^*(\tau)\Big).
\end{equation}
Since $n - m\leq \sum_{j:j\neq i} S_{ij}^*(\tau)\leq n - 1$ for all $\tau\in(0, \mathcal{T}]$ and all $i$,
\begin{equation}
    (n - m)Y_*(\tau) \leq Y(\tau) \leq (n - 1)Y_*(\tau).
    \label{eq:MAapprox}
\end{equation}
Let $\hat{\Lambda}_n(\tau)$ be the Nelson-Aalen estimate of $\Lambda(\tau)$ obtained in a population of size $n$ when who-infects-whom is observed, and let 
\begin{equation}
    \hat{\Lambda}_*(\tau) = \int_0^\tau \frac{\mathbf{1}_{Y_*(u) > 0}}{Y_*(u)}\dif N^*(u).
\end{equation}
Since $n\gg m$, $Y(u) > 0$ if and only if $Y_*(u) > 0$.  Thus,  
\begin{equation}
    \hat{\Lambda}_*(\tau) \leq (n - 1)\hat{\Lambda}_n(\tau) \leq \frac{n - 1}{n - m}\hat{\Lambda}(\tau).
    %\label{hatLambda0}
\end{equation}
by equations~\eqref{eq:hatLambda} and~\eqref{eq:MAapprox}.  Therefore, $\hat{\Lambda}_*(\tau) \rightarrow (n - 1)\hat{\Lambda}_n(\tau)$ for a fixed $m$ as $n\rightarrow\infty$.  Since $\hat{\Lambda}(\tau)$ is a consistent estimator of $\Lambda_n(\tau)$, this implies that $\hat{\Lambda}_*(\tau)$ is a consistent estimator of $\Lambda_*(\tau)$ if we take limits first as $n\rightarrow\infty$ and then as $m\rightarrow\infty$.  Similarly,
\begin{equation}
    \hat{\sigma}^2_*(\tau) = \int_0^\tau \frac{\mathbf{1}_{Y_*(u) > 0}}{Y_*(u)^2}\,dN^*(u).
    %\label{hatsigma0}
\end{equation}
is a consistent estimate of the variance of $\hat{\Lambda}_*(\tau)$.  Using the martingale central limit theorem and a log transformation, we get the asymptotic $1 - \alpha$ confidence limits 
\begin{equation}
    \hat{\Lambda}_*(\tau)\exp\Big(\pm\frac{\hat{\sigma}_*(\tau)}{\hat{\Lambda}_*(\tau)}\Phi^{-1}\big(1 - \frac{\alpha}{2}\big)\Big),
    %\label{hatLambda0CI}
\end{equation}
where we take limits first as $n\rightarrow\infty$ and then as $m\rightarrow\infty$.

When who-infects-whom is not observed, we have the approximate marginal Nelson-Aalen estimate
\begin{equation}
    \widetilde{\Lambda}_*(\tau|\lambda_*) = \int_0^\tau \frac{\mathbf{1}_{Y_*(u) > 0}}{Y_*(u)}\dif\widetilde{N}^*(u|\lambda_*).
    %\label{eq:tildeLambda*}
\end{equation}
Since the probabilities $p_{ij}$ in equation~\eqref{eq:pij} depend on $\lambda^{(k + 1)}(\tau)$ only up to a multiplicative constant, they can be estimated by smoothing the normalized cumulative hazard function $\widetilde{\Lambda}_*(\tau|\lambda_*^{(k)})$.  Thus, $\widetilde{\Lambda}_*(\tau|\lambda_*)$ can replace $\widetilde{\Lambda}(\tau|\lambda)$ in the EM algorithm from Section~\ref{sec:EMalgorithm}.  Let $\widetilde{\Lambda}_*(\tau)$ denote the marginal Nelson-Aalen estimate to which this algorithm converges, let $\widetilde{\lambda}_*(\tau)$ denote the corresponding normalized hazard function, and let $\widetilde{N}^*(\tau) = \widetilde{N}^*\big(\tau\big|\widetilde{\lambda}_*\big)$.  The variance of $\widetilde{\Lambda}_*(\tau)$ is
\begin{equation}
    \widetilde{\sigma}^2_*(\tau) = 2\int_0^\tau \frac{\mathbf{1}_{Y_*(u) > 0}}{Y_*(u)^2}\dif\widetilde{N}^*(u) - \sum_{j=1}^n \Big(\int_0^\tau \frac{\mathbf{1}_{Y_*(u) > 0}}{Y_*(u)}\dif\widetilde{N}^*_{\cdot j}(u)\Big)^2,
    \label{tildesigma0}
\end{equation}
where $\widetilde{N}^*_{\cdot j}(\tau) = \widetilde{N}^*_{\cdot j}\big(\tau\big|\widetilde{\lambda}_*\big)$.  The asymptotic $1 - \alpha$ confidence limits are
\begin{equation}
    \widetilde{\Lambda}_*(\tau)\exp\Big(\pm\frac{\widetilde{\sigma}_*(\tau)}{\widetilde{\Lambda}_*(\tau)}\Phi^{-1}\big(1-\frac{\alpha}{2}\big)\Big).
    %\label{tildeLambda0CI}
\end{equation}

\section{Simulations}
\label{sec:sims}
To evaluate the performance of the methods from Section~\ref{sec:methods}, we conducted simulations of network-based and mass-action epidemics.  In each simulation, we used data on the first 1,000 infections from an epidemic in a population of 100,000 to calculate the Nelson-Aalen and marginal Nelson-Aalen estimates of $\Lambda(\tau)$ and the Kaplan-Meier and marginal Kaplan-Meier estimates of the corresponding survival function.  For each of these estimators, we looked at whether the confidence interval contained the true value of the estimated function at the 5\th, 10\th, 25\th, 50\th, 75\th, 90\th, and 95\th\ percentiles of all possible (i.e., censored and uncensored) contact intervals.  After 1,000 simulations, we calculated the coverage probability for each estimator at each quantile.  For each coverage probability, we calculated an exact 95\% confidence interval.

All models had a latent period of zero and an exponential infectious period with mean one.  The smoothing step of the EM algorithm was done via an inverse-variance weighted cubic smoothing spline with a smoothing parameter chosen by generalized cross-validation \citep{WegmanWright}.  Convergence of the EM algorithm was monitored by calculating the mean of the absolute values of the differences between the current and previous marginal Nelson-Aalen estimates at the 5\th, 10\th, 15\th,\ldots, 85\th, 90\th, and 95\th\ percentiles of the possible contact intervals.  Informally, we call this the ``L1 difference''.  All EM algorithms began by assuming that all possible transmission networks were equally likely, which is equivalent to assuming an exponential contact interval distribution.  We ran the algorithm for a minimum of 5 iterations, and it was continued until an L1 difference less than a specified tolerance was achieved or the 50\th\ iteration was completed.

All epidemic models were written in Python (\texttt{www.python.org}) using the NumPy and SciPy packages (\texttt{www.scipy.org}).  For network-based models, our contact networks were generated using the NetworkX package (\texttt{networkx.lanl.gov}).  Statistical analysis was conducted in R (\texttt{www.r-project.org}) via the RPy2 package (\texttt{rpy.sourceforge.net}).  The code for the models and estimators is available as Online Supplementary Information.

\subsection{Network-based simulations}
In all network-based simulations, the contact network was a Watts-Strogatz small-world network \citep{WattsStrogatz}, which mimics the high clustering and low diameter of real human contact networks.  Starting with a ring of 100,000 nodes, each node was connected to its 10 nearest neighbors.  Each edge was then rewired to a randomly chosen node with probability .1.  Thus, 34.9\% of nodes are connected only to their ten nearest neighbors, 38.7\% are have one long-distance edge, 19.4\% have two long-distance edges, 5.7\% have three long-distance edges, and 1.3\% have four or more long-distance edges.  This network structure gave us high clustering, so infected nodes typically had several possible infectors.  A new contact network was built for each simulation, so the results do not reflect the idiosyncrasies of any particular realization of the network.

We used Weibull(.5, 1), exponential(1), and Weibull(2, 1) contact interval distributions, where Weibull($s, r$) denotes a Weibull distribution with shape parameter $s$ and rate parameter $r$.  The corresponding cumulative hazard functions are:
\begin{align}
    \Lambda(\tau) &= \sqrt{\tau} \text{ for the Weibull(.5, 1) distribution}\\
    \Lambda(\tau) &= \tau \text{ for the exponential(1) distribution, and}\\
    \Lambda(\tau) &= \tau^2 \text{ for the Weibull(2, 1) distribution}.
\end{align}
Note that the exponential(1) distribution is the same as a Weibull(1, 1) distribution.  The hazard of infectious contact decreases throughout the infectious period in the Weibull(.5, 1) case, remains constant in the exponential(1) case, and increases throughout the infectious period in the Weibull(2, 1) case.  

The L1 tolerance for the EM algorithm was set to .0005.  This was approximately the minimum tolerance consistently achieved by the EM algorithm before the L1 differences became a chaotic sequence of small numbers. 

\begin{figure}
    \includegraphics[width = \textwidth]{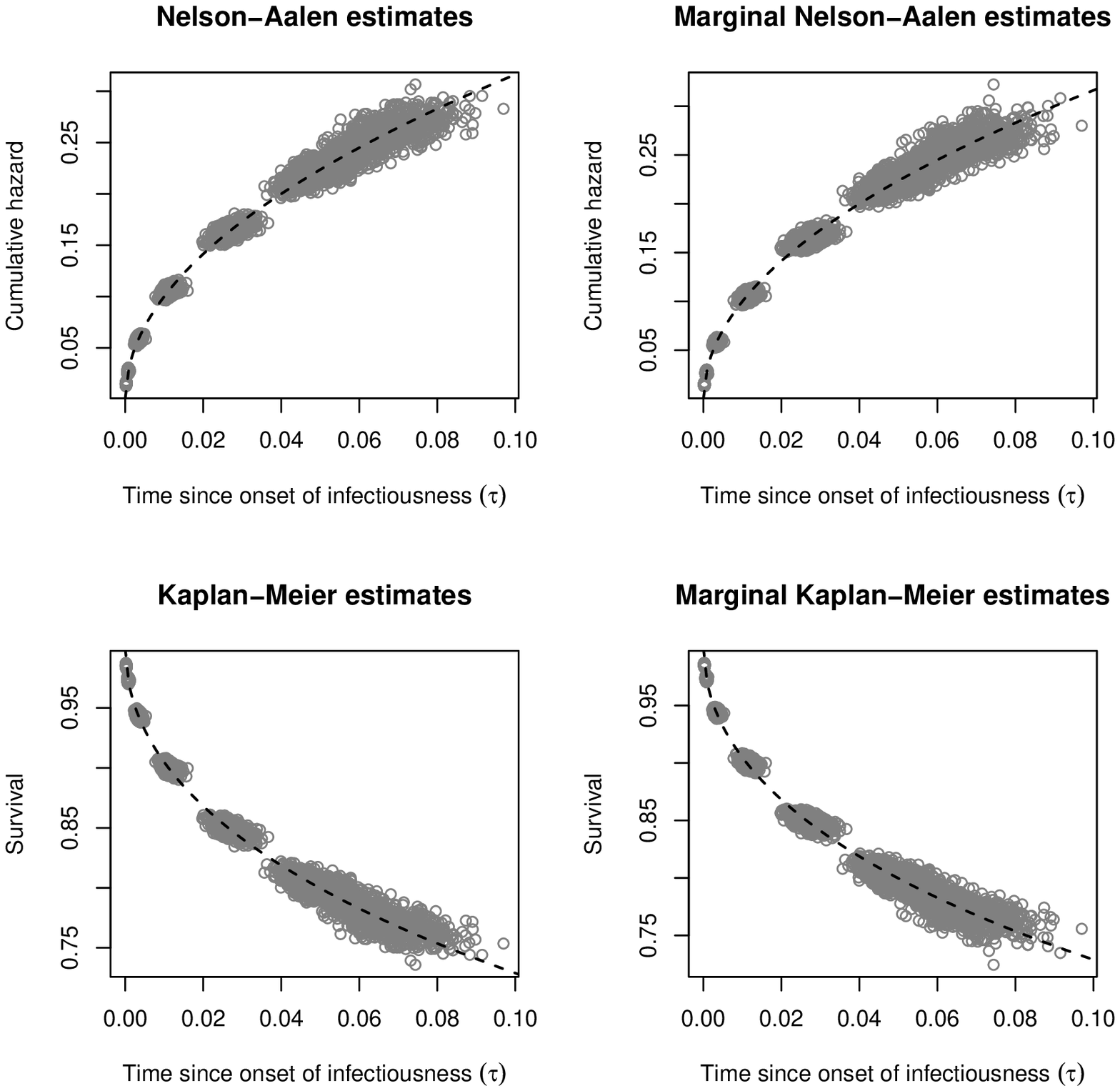}
    \caption{Estimates of the cumulative hazard and survival functions for network-based models with a Weibull(.5, 1) contact interval distribution, with (left) and without (right) data on who-infected-whom.  The true cumulative hazard function (top) and survival function (bottom) are indicated with dashed lines.  The point clouds are point estimates at the 5\th, 10\th, 25\th, 50\th, 75\th, 90\th, and 95\th\ percentiles of the possible contact intervals in each of 1,000 simulations.}
    \label{fig:wsW.5}
\end{figure}

\begin{figure}
    \includegraphics[width = \textwidth]{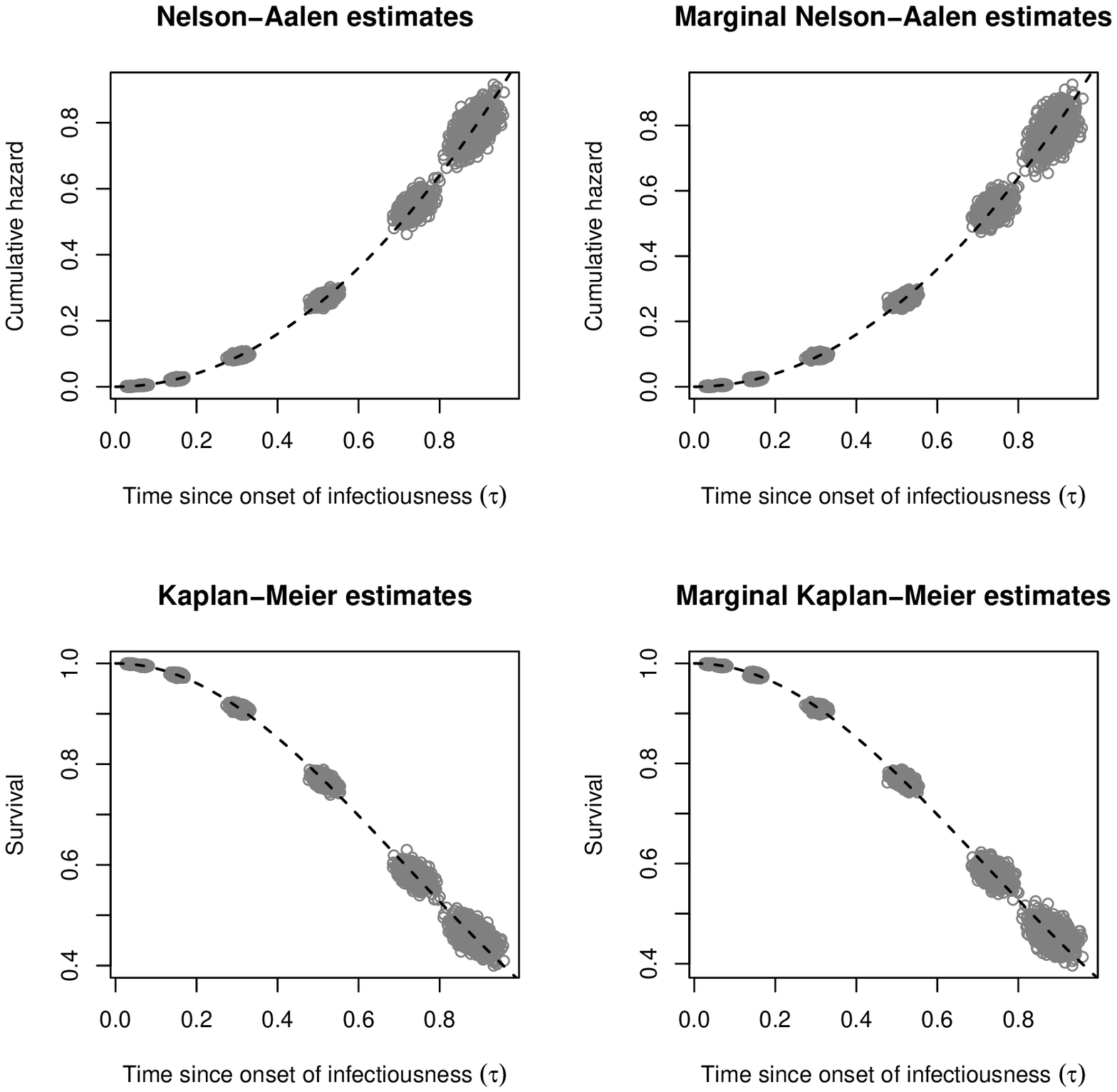}
    \caption{Estimates of the cumulative hazard and survival functions for network-based models with a Weibull(2, 1) contact interval distribution, with (left) and without (right) data on who-infected-whom.  The true cumulative hazard function (top) and survival function (bottom) are indicated with dashed lines.  The point clouds are point estimates at the 5\th, 10\th, 25\th, 50\th, 75\th, 90\th, and 95\th\ percentiles of the possible contact intervals in each of 1,000 simulations.}
    \label{fig:wsW2}
\end{figure}

\paragraph{Results}
The EM algorithm for the marginal Nelson-Aalen estimate converged easily in all simulations.  For models with exponential(1) contact intervals, the desired tolerance was achieved within 5 iterations for all models.  For Weibull(.5, 1) contact intervals, the maximum number of iterations was 6.  For the Weibull(2, 1) contact intervals, the maximum number was 8.  Table~\ref{tab:netsims} shows the coverage probabilities and exact 95\% confidence intervals for the Nelson-Aalen, marginal Nelson-Aalen, Kaplan-Meier, and marginal Kaplan-Meier estimators in models with Weibull(.5, 1) and Weibull(2, 1) contact interval distributions.  Coverage probabilities for the Nelson-Aalen and Kaplan-Meier estimators was close to .95 across the range of available data for both models.  Coverage probabilities for the marginal Nelson-Aalen and Kaplan-Meier estimators was slightly lower but above .90 in all cases.  Coverage probabilities for the model with exponential contact intervals were above .93 for all estimators and data quantiles (not shown).  Figures \ref{fig:wsW.5} and \ref{fig:wsW2} show good agreement between the estimated and true cumulative hazard and survival functions for the contact interval distribution across the range of available data.  The true cumulative hazard function (top) and survival function (bottom) are indicated by dashed lines.  The Nelson-Aalen and Kaplan-Meier estimates (left) use information on who-infected-whom.  The marginal Nelson-Aalen and Kaplan-Meier estimates (right) do not use this information.  The point clouds are the point estimates at the 5\th, 10\th, 25\th, 50\th, 75\th, 90\th, and 95\th\ percentiles of the possible contact intervals in each of 1,000 simulations.  The clouds for the 5\th\ and 10\th\ and the 90\th\ and 95\th\ percentiles often run together, leaving five separate clouds of points.

\subsection{Mass-action simulations}
For mass-action models, we used Weibull(.5, 5), exponential(2), and Weibull(2, 1) normalized contact interval distributions.  The corresponding cumulative hazard functions are:
\begin{align}
    \Lambda_*(\tau) &= \sqrt{5\tau} \text{ for the Weibull(.5, 5) distribution}\\
    \Lambda_*(\tau) &= \tau \text{ for the exponential(1) distribution, and}\\
    \Lambda_*(\tau) &= \tau^2 \text{ for the Weibull(2, 1) distribution}.
\end{align}
In a mass-action model, a Weibull(.5, 1) distribution for the normalized contact interval distribution produces $R_0 = .89$, so no epidemics occur.  Changing the rate parameter to 5 produces $R_0 = 1.98$, which is very close to the $R_0 = 2$ of the other two models.  These are $R_0$ estimates assuming a network with no clustering \citep{Kenah4}, which places an upper bound on the true $R_0$ in networks with clustering.  The hazard of infectious contact decreases throughout the infectious period in the Weibull(.5, 5) case, remains constant in the exponential(1) case, and increases throughout the infectious period in the Weibull(2, 1) case.

The L1 tolerance for the EM algorithm was set to .005.  This was approximately the minimum tolerance consistently achieved by the EM algorithm before the L1 differences became a chaotic sequence of small numbers. 

\begin{figure}
    \includegraphics[width = \textwidth]{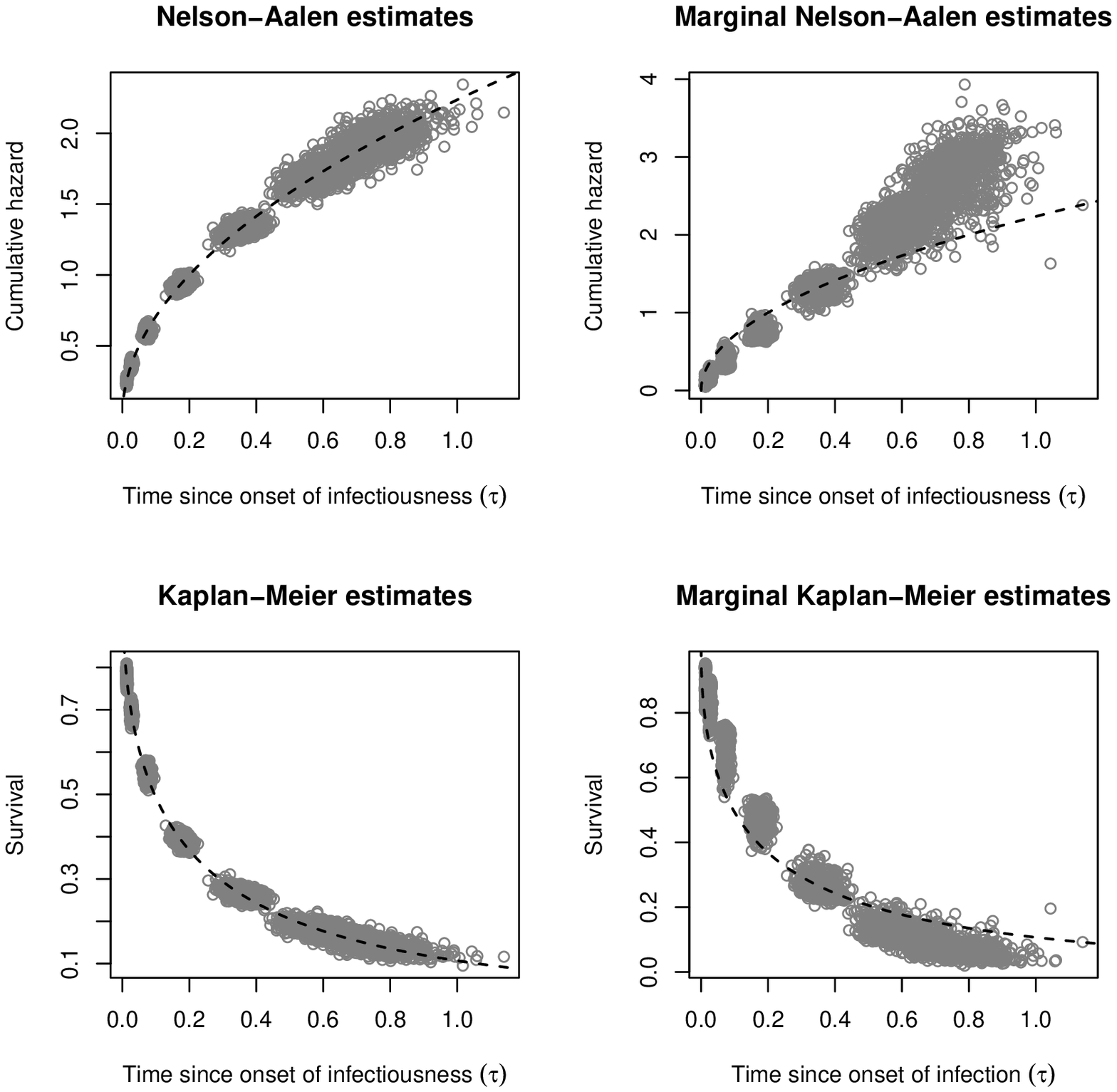}
    \caption{Estimates of the cumulative hazard and survival functions for mass-action models with a Weibull(.5, 5) normalized contact interval distribution using the methods from Section~\ref{sec:massaction}, with (left) and without (right) data on who-infected-whom.  The true cumulative hazard function (top) and survival function (bottom) are indicated with dashed lines.  The point clouds are point estimates at the 5\th, 10\th, 25\th, 50\th, 75\th, 90\th, and 95\th\ percentiles of the possible contact intervals in each of 1,000 simulations.  Note the deviation of the marginal Nelson-Aalen and Kaplan-Meier estimates from the true functions.}
    \label{fig:maW.5}
\end{figure}

\begin{figure}
    \includegraphics[width = \textwidth]{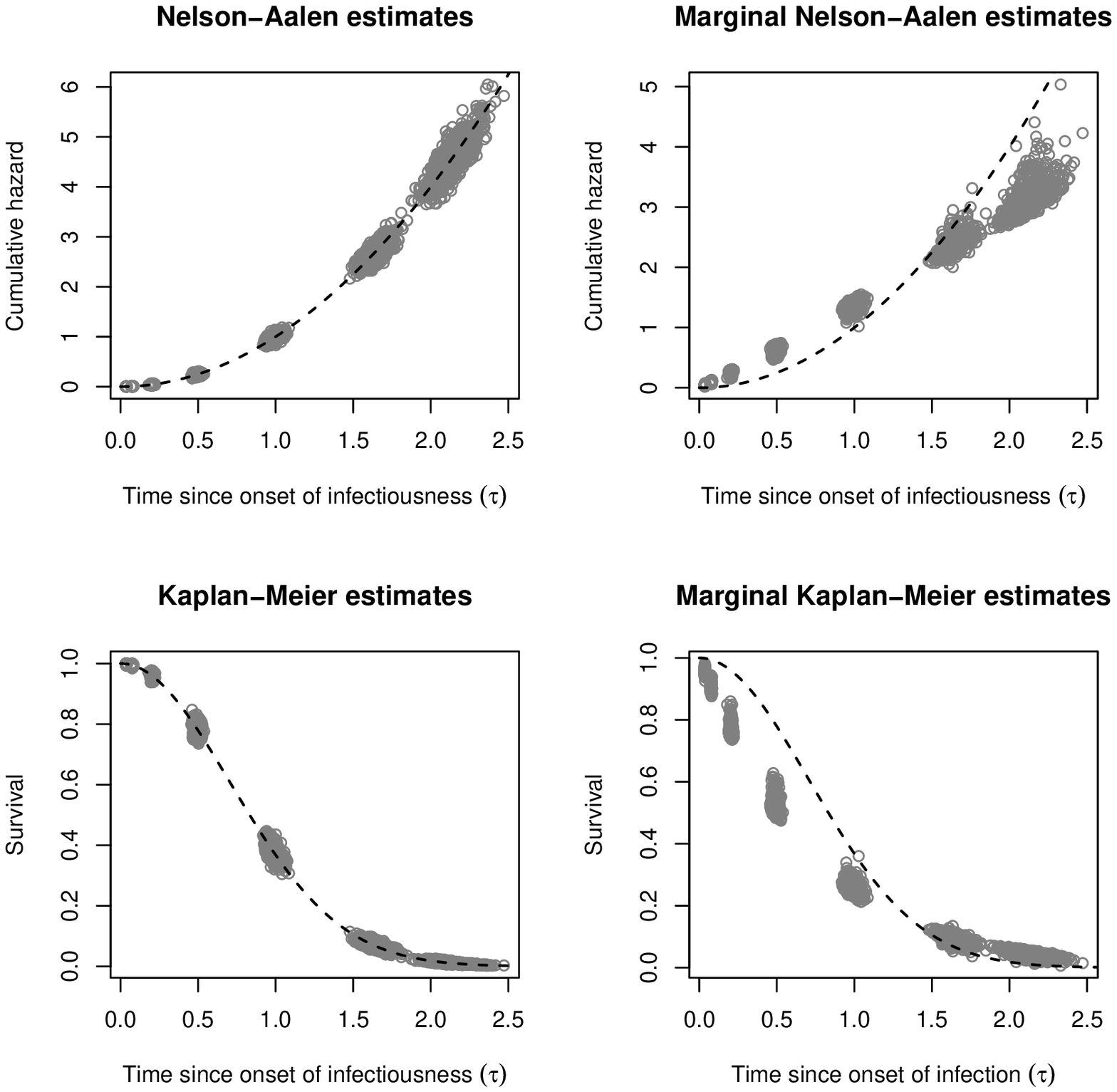}
    \caption{Estimates of the cumulative hazard and survival functions for mass-action models with a Weibull(2, 1) normalized contact interval distribution using the methods from Section~\ref{sec:massaction}, with (left) and without (right) data on who-infected-whom.  The true cumulative hazard function (top) and survival function (bottom) are indicated with dashed lines.  The point clouds are point estimates at the 5\th, 10\th, 25\th, 50\th, 75\th, 90\th, and 95\th\ percentiles of the possible contact intervals in each of 1,000 simulations.  Note the deviation of the marginal Nelson-Aalen and Kaplan-Meier estimates from the true functions.}
    \label{fig:maW2}
\end{figure}

\paragraph{Results}
The EM algorithm for the marginal Nelson-Aalen estimate took longer to converge for mass-action models than for network-based models, despite the greater tolerance.  For models with exponential contact intervals, the maximum number of iterations required was 19.  For Weibull(2, 1) contact intervals, the maximum was 29.  For Weibull(.5, 5) contact intervals, the maximum was 32.  Table~\ref{tab:masims} shows coverage probabilities and exact 95\% confidence intervals for the Nelson-Aalen, marginal Nelson-Aalen, Kaplan-Meier, and marginal Kaplan-Meier estimators in models with Weibull(.5, 5) and Weibull(2, 1) normalized contact interval distributions.  Coverage probabilities for the Nelson-Aalen and Kaplan-Meier estimators were close to .95 across the range of available data for both models.  In contrast, the marginal Nelson-Aalen and Kaplan-Meier estimators were spectacular failures.  In the model with exponential normalized contact interval distributions, coverage probabilities for Nelson-Aalen and Kaplan-Meier estimators were all above .93 and coverage probabilities for the marginal Nelson-Aalen and Kaplan-Meier estimators were all above .99.  Figures \ref{fig:maW.5} and \ref{fig:maW2} show good agreement between the estimated and true survival and cumulative hazard functions when who-infected-whom is observed (left) but poor agreement when who-infected-whom is not observed (right).  The dashed lines and clouds of points have the same interpretation as in Figures~\ref{fig:wsW.5} and~\ref{fig:wsW2}.

The number of possible infectors for each infection is much, much larger in a mass-action model than in a network-based model.  It seems likely that the EM algorithm cannot usefully assign probabilities to the possible infectors of each infected person because the asymptotic approximation to the marginal Nelson-Aalen estimate is not sufficiently precise and because of limited numerical precision.  In the case of exponential contact intervals, the initial step of the EM algorithm correctly guesses that all possible infectors are equally likely to have been the source of infection, so the EM algorithm converges.  The large coverage probabilities suggest that the variance approximation from equation~\eqref{tildesigma0} overestimates the true variance.

\section{Data analysis: Pandemic influenza A(H1N1) in Los Angeles County, 2009}
\label{sec:data}
In this section, we use the marginal Nelson-Aalen estimator to analyze pandemic influenza A(H1N1) (pH1N1) surveillance data collected by the Los Angeles County Department of Public Health between April 22 and May 19, 2009.  The data was collected according to the following protocol \citep{LAh1n1}:
\begin{enumerate}
    \item Individuals who presented to healthcare providers or the county health department with acute febrile respiratory illness (AFRI, defined as fever $\geq 100^{\circ}\mathrm{F}$ plus cough, sore throat, or runny nose) had nasopharyngeal swabs and aspirates tested for pandemic influenza A(H1N1).  Those who tested positive are the \textit{index cases}.
    \item The index cases were interviewed by telephone and asked to report AFRI episodes among household contacts, including the dates of illness onset.  Additional AFRI episodes among household contacts were ascertained during follow-up interviews 14 days after the illness onset of the index case.  All cases of AFRI in the household after 10 days prior to symptom onset in the index case were assumed to be pH1N1 cases.  The earliest pH1N1 case in each household is the \textit{primary case}; this is usually (but not always) the index case.
\end{enumerate}
There were 58 households with a total of 299 members.  In these households, there were 62 primary cases (four households had co-primary cases) and 35 secondary cases.  In 51 of the 58 households, the index case was a primary case.  This is a good example of household surveillance data from the early stages of an epidemic.  

As in the simulations, our goal is to estimate the cumulative hazard function of the contact interval distribution.  We compare the marginal Nelson-Aalen estimate with parametric estimates obtained using the methods from \citet{Kenah4}.  We use the corresponding marginal Kaplan-Meier estimate to estimate the probability that an infected person makes infectious contact with a given household member during his or her infectious period, which we call the \textit{household infectious contact probability}.

Our natural history assumptions are adapted from \citet{YangH1N1}.  In the primary analysis, we assumed an incubation period of 2 days, a latent period of 0 days, and an infectious period of 6 days.  This means that a person $i$ with onset of symptoms on day $t_i^\text{sym}$ is infected on day $t_i = t_i^\text{sym} - 2$, has onset of infectiousness on day $t_i$, and recovers from infectiousness on day $t_i + 6$.  In discrete time, the first day on which $i$ is able to infect other persons is the day after his or her onset of infectiousness, which is $t_i + 1 = t_i^\text{sym} - 1$ under our assumptions.  If day 0 is the infection time, then we have the first secondary transmissions on day 1, the onset of symptoms on day 2, and recovery on day 6.  In a sensitivity analysis, we vary the incubation period from 1 day to 3 days, the infectious period from 5 days to 7 days, and the latent period from 0 days to 1 day.  

\subsection{Results}
\begin{figure}
    \includegraphics[width = \textwidth]{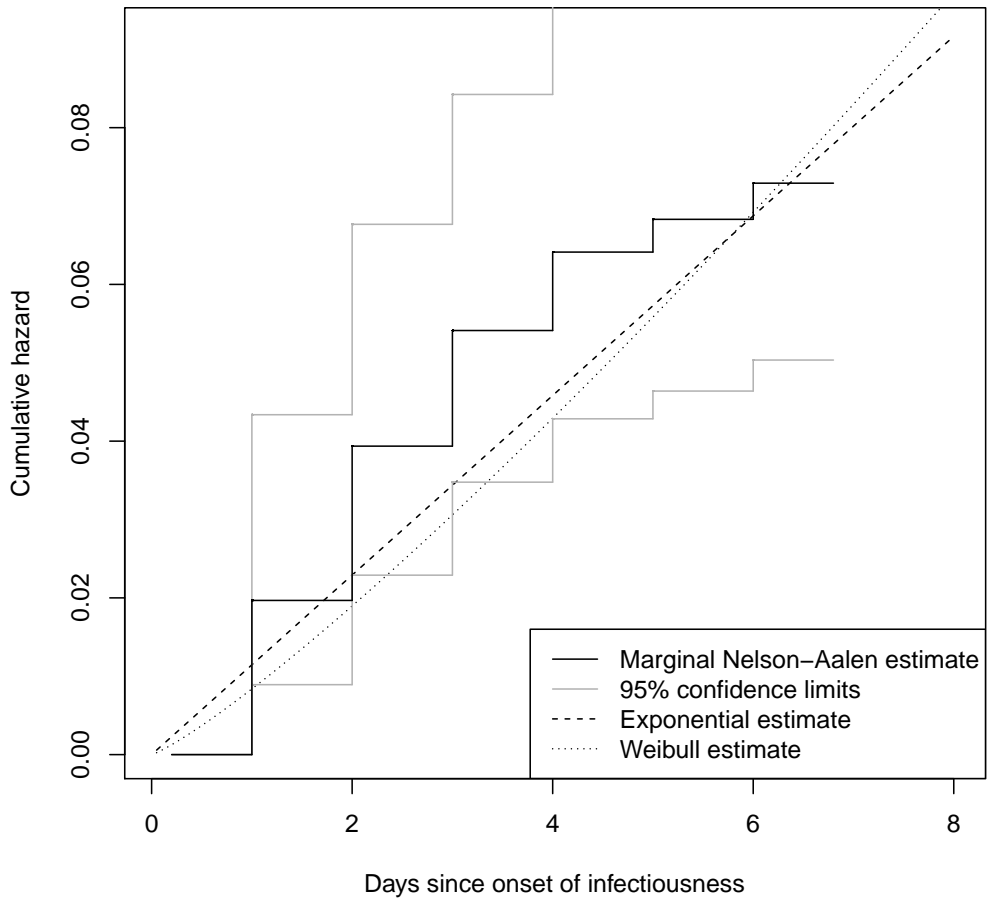}
    \caption{A marginal Nelson-Aalen estimate of the contact interval cumulative hazard function based on the Los Angeles County household data.  For comparison, we have parametric estimates of the cumulative hazard function assuming exponential and Weibull contact interval distributions.  The assumed latent period for all analyses was 2 days, so symptom onset occurs on day 2 after the onset of infectiousness.  Note the large jumps on days 1, 2, and 3 following the onset of infectiousness and the small jumps on days 5 and 6.}
    \label{fig:LAestimates}
\end{figure}

Figure~\ref{fig:LAestimates} shows the marginal Nelson-Aalen estimate of the cumulative hazard function of the contact interval distribution, along with approximate 95\% confidence limits and parametric estimates assuming exponential and Weibull contact interval distributions.  A cumulative hazard estimate assuming a gamma contact interval distribution (not shown) was almost exactly the same as that assuming a Weibull distribution.  

The exponential, Weibull, and marginal Nelson-Aalen point estimates of the cumulative hazard at 6 days post-infection are .0687, .0692, and .0729, respectively.  The corresponding survival probabilities are .9334, .9331, and .9296.  Thus, all three estimates indicate a household infectious contact probability of .07 over the course of the infectious period.  The approximate 95\% confidence interval for the marginal Nelson-Aalen estimate is (.0503, .1056).  The corresponding 95\% confidence interval for the marginal Kaplan-Meier estimate of the survival probability is (.8997, .9509).  Therefore, our nonparametric estimate of the household infectious contact probability is .07 (.05, .10).  

While the nonparametric and parametric estimates agree on the household infectious contact probability, they differ in their estimates of distribution of infectiousness over time.  The exponential estimate inherently predicts a constant hazard of infectious contact over the entire infectious period.  The Weibull estimate also predicts little variation in the hazard of infectious contact over the infectious period, with slightly lower infectiousness near the beginning and slightly higher infectiousness near the end.  The marginal Nelson-Aalen curve has much larger jumps on days 1, 2, and 3 following infection than on days 5 and 6, which places the highest infectiousness on the day prior to, the day of, and the day after the onset of symptoms.  According to this estimate, only 12\% of infectious contacts occur $>2$ days after the onset of symptoms and only 6\% occur $>3$ days after symptom onset.    

\subsection{Sensitivity analysis}
\begin{figure}
    \includegraphics[width = \textwidth]{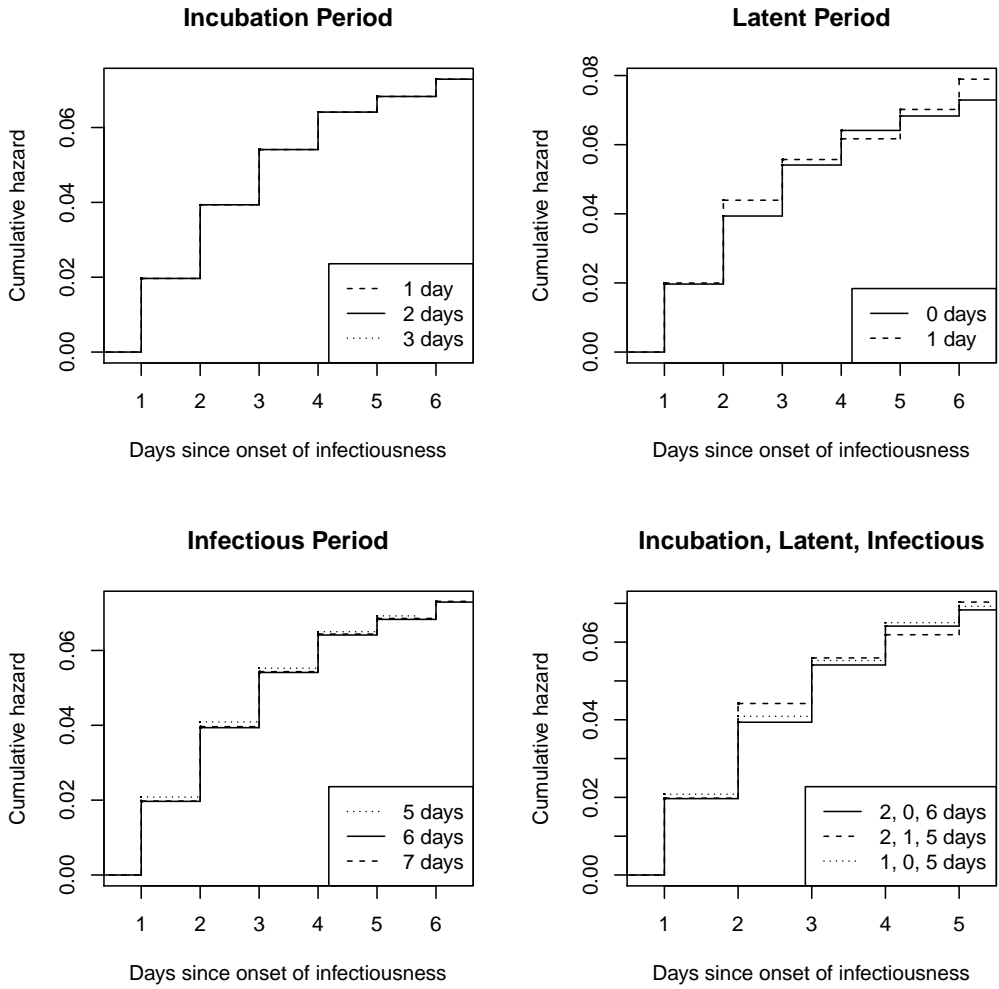}
    \caption{Sensitivity analyses for the nonparametric analysis of the LA county household data.  There is little sensitivity to the assumed incubation and infectious periods.  There is greater sensitivity to the assumed latent period, but there is good evidence for transmission on the day prior to symptom onset and little evidence of significant transmission $\geq 2$ days prior to symptom onset \citep{CauchemezH1N1, DonnellyH1N1}.}
    \label{fig:LAsensitivity}
\end{figure}

Figure~\ref{fig:LAsensitivity} shows the results of a sensitivity analysis in which we vary the assumed (a) incubation period, (b) latent period, and (c) infectious period.  In panel (d), we vary all three of these intervals simultaneously to generate the greatest possible variation.  The results are insensitive to the incubation period and infectious period but sensitive to the latent period.  However, there is good evidence that little or no transmission occurs two or more days prior to the onset of symptoms \citep{DonnellyH1N1}.

\subsection{Previous pH1N1 household data analyses}
Our nonparametric estimate of .07 (.05, .10) for the household infectious contact probability is very similar to the .06 (.03, .11) estimated using data collected by Public Health--Seattle and King County during April-May, 2009 \citep{SugimotoH1N1}.  This estimate was obtained using a chain-binomial model to estimate the probability of infectious contact with a given household member on each day of the infectious period and then calculating the probability that infectious contact occurs at least once, making it directly comparable with our estimate.  The relationship between contact intervals and chain binomial models is explained in Appendix~\ref{app:chainbinom}.

It is more difficult to compare our estimated household infectious contact probability with estimates of the household secondary attack rate (SAR), which are usually obtained by calculating the proportion of all susceptible members of study households who have the onset of a given set of symptoms within a given time period relative to symptom onset in the index case.  Among these estimates for pH1N1 are: .145 (.129, .164) for influenza-like illness (ILI) within 7 days after the index case symptom onset \citep{CarcioneH1N1}, .13 for acute respiratory infection (ARI) and .9 for ILI within 1-9 days after the index case symptom onset \citep{MorganH1N1}, .113 (.088, .137) for ILI up to 23 days after index case symptom onset \citep{FranceH1N1}, and .13 for ARI and .10 for ILI 7 days before or after the index case symptom onset \citep{CauchemezH1N1}.  We did the following simulations to see if our estimated household infectious contact probability of .07 is consistent with these estimates: 
\begin{enumerate}
    \item For each of the 58 households in the LA data, we simulated an epidemic using its observed number of index cases and an household infectious contact probability of .07.  
    \item After the epidemics in all households completed, we calculated the proportion of all household contacts who were infected.  
\end{enumerate}
Repeating this $10,000$ times, we got a mean household SAR of .12 with a bootstrap percentile 95\% confidence interval of (.06, .19), which places us solidly within the range of previous household SAR estimates.

Our nonparametric estimate of the distribution of infectiousness over time is consistent with estimates of the serial interval distribution for pH1N1 by \citet{DonnellyH1N1}, who found high infectiousness on days 0, 1, and 2 following symptom onset with only 18\% of transmission occurring $>2$ days after the onset of symptoms and only 5\% occurring $>3$ days after symptom onset.  A similar pattern of high infectiousness early in the infectious period was found by \citet{CauchemezH1N1}, who also used a serial interval distribution.  Though theoretical and practical problems with the serial interval distribution were noted in Section~\ref{sec:intro}, the similarity between our results and those of \citet{DonnellyH1N1} and \citet{CauchemezH1N1} suggests that the marginal Nelson-Aalen estimate captured an important feature of pH1N1 transmission that was missed by the parametric contact interval distribution estimates.

\section{Discussion}
\label{sec:discussion}
The marginal Nelson-Aalen estimator suffers many of the same limitations as the parametric estimators in \citet{Kenah4} and some of its own.  In roughly decreasing order of difficulty, these include:  
\begin{enumerate}
    \item \label{lim:diseases} The SEIR framework is limited to acute, immunizing infectious diseases that spread person-to-person, so our methods do not apply directly to tuberculosis, HIV/AIDS, many foodborne and waterborne diseases, pneumococcal and meningococcal diseases, and other infectious diseases of major public health importance.  
    \item \label{lim:ind} We have assumed that contact intervals are independent of infectious periods, which is unrealistic.  Both are affected by the interaction between the host and the pathogen.  
    \item \label{lim:data} We have assumed that infection times, times of onset of infectiousness, and times of recovery from infectiousness are observed, which is unrealistic.  Usually, only symptom onset times are observed.  In the analysis of the Los Angeles County household data in Section~\ref{sec:data}, we were forced to assume constant latent, incubation, and infectious periods because we had only symptom onset times.  The implicit assumption of constant latent, incubation, and infectious periods in methods based on generation and serial intervals was one of our main criticisms in Section~\ref{sec:intro}.  Unfortunately, our analysis merely made these same assumptions explicit.
    \item \label{lim:homogeneity} We assumed that the contact interval distribution is identically distributed for all ordered pairs $ij$ such that $C_{ij} = 1$, which is unrealistic.  For instance, age was found to have an effect on pH1N1 transmission in most of the papers cited in Section~\ref{sec:data} \citep{CauchemezH1N1, FranceH1N1, MorganH1N1, CarcioneH1N1, SugimotoH1N1}.  The effects of covariates on the transmission of disease is a central concern of infectious disease epidemiology. 
    \item \label{lim:smoothing} There are several technical issues that need further exploration.  A more rigorous theoretical analysis of convergence of the marginal Nelson-Aalen estimator, including the smoothing step, is needed to better understand the conditions under which it will work.  Convergence criteria need to be more carefully examined, confidence intervals for small samples need to be developed, and more general stopping times for the end of observation need to be considered. 
\end{enumerate}
Nonetheless, survival analysis based on contact intervals is a powerful framework for the development of statistical methods in infectious disease epidemiology.  These methods have important theoretical and practical advantages over methods based on generation or serial intervals, including greater flexibility in the choice of an underlying transmission model, greater clarity about data requirements and analytical assumptions, the exploitation of information contributed by uninfected person-time, and validity throughout the course of an epidemic.  The theory and methods of survival analysis offer many possibilities for overcoming the limitations listed above.  

The arguments for convergence of the Nelson-Aalen estimator given in Section~\ref{sec:methods} are largely heuristic, but the theoretical analysis required to address limitation~(\ref{lim:smoothing}) is complicated by the smoothing step.  In simulations, the performance of the marginal Nelson-Aalen estimator did not appear sensitive to the smoothing method.  However, a smoother designed specifically for hazard functions, such as that of \citet{MullerWang}, may improve performance under certain conditions.  Another technical limitation is that the variance estimates based on equation~\eqref{eq:condvar} treat the estimated transmission network probabilities as known, ignoring a component of the variance that may be important in small samples.  These and other technical issues need to be resolved with more rigorous theoretical analysis and further simulation studies.

Limitation~(\ref{lim:homogeneity}) can be addressed by incorporating the marginal Nelson-Aalen estimate into a semiparametric regression model for the effects of covariates on the hazard of infectious contact.  Suppose the hazard of infectious contact from person $i$ to person $j$ at infectiousness age $\tau$ of $i$ is
\begin{equation}
    \lambda_{ij}(\tau) = \lambda_0(\tau)\exp(\beta^\text{T} X_{ij}),
\end{equation}
where $\beta$ is a parameter vector and $X_{ij}$ is a vector of infectiousness covariates for $i$, susceptibility covariates for $j$, and pairwise covariates for $ij$.  The methods of Section~\ref{sec:methods} could be extended to nonparametrically estimate $\lambda_0(\tau)$ by adding a step to the EM algorithm that maximized the expected log likelihood over $\beta$.  This would yield a semiparametric estimator of the effects of covariates on infectiousness and susceptibility that is similar in spirit to the Cox proportional hazards model.  In infectious disease epidemiology, these covariate effects are usually a higher priority than the distribution of infectiousness over time, so the marginal Nelson-Aalen estimator may find its most important applications behind the scenes of this regression model.

Limitation~(\ref{lim:data}) could be addressed by adopting a semiparametric Bayesian framework \citep{SinhaDey} or using a profile sampler \citep{LeeKosorokFine}.  A semiparametric Bayesian approach would use a prior process instead of a marginal Nelson-Aalen estimator.  A profile sampler would take advantage of the fact that the marginal Nelson-Aalen estimate is an efficient numerical method of obtaining a likelihood maximized over all possible contact interval distributions for a given set of infection times, infectiousness onset times, and recovery times.  

Limitation~(\ref{lim:ind}) could be addressed by using multivariate survival methods to estimate the joint distribution of the contact interval and the infectious period, and limitation~(\ref{lim:diseases}) could be addressed by allowing individuals to experience multiple infection events or infection events of different types (e.g., new infection, carriage, relapse).

Survival analysis based on contact intervals is a promising approach to the statistical analysis of infectious disease data.  The marginal Nelson-Aalen estimator extends a beautiful and powerful set of nonparametric methods from standard survival analysis to infectious disease epidemiology.  The simulations in Section~\ref{sec:sims} show that these methods are reliable, and the data analysis in Section~\ref{sec:data} shows that the nonparametric methods in this paper are an important extension of the parametric methods in \citet{Kenah4}.  Each of the ingredients of this approach have been applied to infectious diseases before: counting processes and martingales by \citet{Becker} and \citet{Rhodes}, the EM algorithm by \citet{BeckerEM}, and summation over possible transmission networks by \citet{WallingaTeunis}.  The marginal Nelson-Aalen estimator combines these in a novel and useful way, correcting approaches based on generation and serial intervals, generalizing the chain binomial model, and pointing the way to more flexible and practical statistical methods for infectious disease epidemiology.

\section*{Acknowledgements}
This work benefited greatly from the comments of M.\ Elizabeth Halloran, Ira M. Longini, Jr., Yang Yang, Jonathan D.\ Sugimoto, and participants in the Dhaka University Statistics Department Alumni Association International Conference on the Theory and Application of Statistics (December 26-28, 2010 in Dhaka, Bangladesh), especially Md.\ Abu Yushuf Sharker.  Brit Oiulfstad, Dee Ann Bagwell, Brandon Dean, Laurene Mascola, and Elizabeth Bancroft of the Los Angeles County Department of Public Health generously allowed the use of their data in Section~\ref{sec:data}.  Office space, computing facilities, and administrative support were provided by the Fred Hutchinson Cancer Research Center.  This research was supported by National Institute of General Medical Sciences (NIGMS) grant F32GM085945.  The content is solely the responsibility of the author and does not necessarily represent the official views of NIGMS or the National Institutes of Health.

\bibliographystyle{chicago}
\bibliography{npCIestimation}

\appendix
\section{Partially-observed transmission networks}
\label{app:partialv}
In a partially-observed transmission network, we observe the infectors of a subset of all infected persons.  By taking $\mathcal{V}_j = v_j$ for all $j$ such that the infector is observed, equations~\eqref{eq:tildeLambda1} through~\eqref{eq:tildeLambda} define a marginal Nelson-Aalen estimator that takes this additional information into account.  For each $j$ with an observed $v_j$, $\widetilde{N}^*_{\cdot j}(\tau) = N_{v_j j}(\tau)$.  The variance estimate in equation~\eqref{eq:tildesigma} can be used to obtain confidence intervals via equation~\eqref{eq:tildeLambdaCI}.  

\section{Contact intervals and chain binomial models}
\label{app:chainbinom}
In this appendix, we show that the methods of Section~\ref{sec:methods} reproduce classical chain-binomial models when the contact interval distribution is discrete.  In the case of a continuous contact interval distribution \citep{Kenah3, Kenah4}, the likelihood when who-infected-whom is not observed is
\begin{equation}
    L(S) \;= \prod_{j:0 < t_j\leq T} \big[\sum_{i\in\mathcal{V}_j} \lambda(t_j - t_i - \varepsilon_i)\big]\prod_{i:t_i < t_j} S\big(\iota_i\wedge [t_j - t_i - \varepsilon_i]\big)^{C_{ij}},
\end{equation}
where $\lambda(\tau) = \frac{S'(\tau)}{S(\tau)}$.  If we take $\lambda(\tau) = \frac{S(\tau^-) - S(\tau)}{S(\tau)}$, this is equivalent to the integrated likelihood maximized by the EM algorithm in Section~\ref{sec:methods}.

When the contact interval and latent period distributions are discrete, we must account for the possibility of tied infectious contact times.  Instead of each infected person $j$ having a single infector, the source of his or her infection can be any nonempty subset $V_j\subseteq\mathcal{V}_j$.  Therefore, the hazard term for $j$ in the likelihood becomes
\begin{equation}
    \sum_{\emptyset\neq V_j\subseteq\mathcal{V}_j} \prod_{i\in V_j} \lambda(t_j - t_i - \varepsilon_i)\prod_{i\not\in V_j} \big[1 - \lambda(t_j - t_i - \varepsilon_i)\big],
    \label{eq:discretehaz}
\end{equation}
where 
\begin{equation}
    \lambda(\tau) = \frac{S(\tau^-) - S(\tau)}{S(\tau^-)}
\end{equation}
is the conditional probability of infectious contact at $\tau$ given no infectious contact before $\tau$.  Equation~\eqref{eq:discretehaz} contains all terms in 
\begin{equation}
    \prod_{i\in\mathcal{V}_j} \big[1 - \lambda(t_j - t_i - \varepsilon_i)\big] + \lambda(t_j - t_i - \varepsilon_i) = 1
\end{equation}
except
\begin{equation}
    \prod_{i\in\mathcal{V}_j} \big[1 - \lambda(t_j - t_i - \varepsilon_i)\big],
\end{equation}
so the likelihood for a discrete contact interval distribution can be written
\begin{equation}
    L(S) \;= \prod_{j:0 < t_j\leq T} \Big(1 - \prod_{i\in\mathcal{V}_j} \big[1 - \lambda(t_j - t_i - \varepsilon_i)\big]\Big)\prod_{i:t_i < t_j} S\big(\iota_i\wedge [t_j - t_i - \varepsilon_i]\big)^{C_{ij}}.
    \label{eq:chainbinom}
\end{equation}
Since $1 - \lambda(\tau)$ can be interpreted as an infection escape probability, the likelihood in equation~\eqref{eq:chainbinom} is precisely the same as that of a chain-binomial model \citep{RampeyLongini}.  Therefore, chain-binomial models can be defined in terms of contact interval distributions, and the methods of this paper can be seen as an extension of chain-binomial models to continuous time.

\clearpage
\begin{table}
    \caption{\label{tab:netsims} 95\% confidence interval coverage probabilities in network-based simulations.}
    \centering
    \fbox{%
    \begin{tabular}{r|rr|rr}
	\textbf{Data} & \multicolumn{4}{c}{\textbf{Contact interval distribution}}\\
	\textbf{percentile} & \multicolumn{2}{c|}{\textbf{Weibull(.5, 1)}} & \multicolumn{2}{c}{\textbf{Weibull(2, 1)}}\\
	& Nelson-Aalen  & marginal NA  & Nelson-Aalen  & marginal NA \\
	5 & .949 (.933, .962) & .946 (.930, .959) & .960 (.946, .971) & .903 (.883, .921)\\
	10 & .954 (.939, .966) & .952 (.937, .964) & .963 (.949, .974) & .901 (.881, .919)\\
	25 & .957 (.943, .969) & .947 (.931, .960) & .952 (.937, .964) & .929 (.911, .944)\\
	50 & .952 (.937, .964) & .940 (.923, .954) & .944 (.928, .957) & .924 (.906, .940)\\
	75 & .948 (.932, .961) & .944 (.928, .957) & .958 (.944, .970) & .958 (.944, .970)\\
	90 & .935 (.918, .949) & .944 (.928, .957) & .956 (.941, .968) & .939 (.922, .953)\\
	95 & .934 (.917, .949) & .938 (.921, .952) & .942 (.926, .956) & .935 (.918, .949)\\[5pt]
	& Kaplan-Meier  & marginal KM  & Kaplan-Meier  & marginal KM\\
	5 & .949 (.933, .962) & .946 (.930, .959) & .960 (.946, .971) & .903 (.883, .921)\\
	10 & .954 (.939, .966) & .952 (.937, .964) & .964 (.951, .975) & .901 (.881, .919)\\
	25 & .957 (.943, .969) & .947 (.931, .960) & .952 (.937, .964) & .929 (.911, .944)\\
	50 & .952 (.937, .964) & .940 (.923, .954) & .944 (.928, .957) & .925 (.907, .941)\\
	75 & .948 (.932, .961) & .945 (.929, .958) & .960 (.946, .971) & .957 (.942, .969)\\
	90 & .934 (.917, .949) & .944 (.928, .957) & .955 (.940, .967) & .940 (.923, .954)\\
	95 & .933 (.916, .948) & .938 (.921, .952) & .943 (.927, .957) & .936 (.919, .950)
    \end{tabular}
    }
\end{table}

\begin{table}
    \caption{\label{tab:masims} 95\% confidence interval coverage probabilities for mass-action simulations.}
    \centering
    \fbox{%
    \begin{tabular}{r|rr|rr}
	Data & \multicolumn{4}{c}{Scaled contact interval distribution}\\
	percentile & \multicolumn{2}{c|}{Weibull(.5, 5)} & \multicolumn{2}{c}{Weibull(2, 1)}\\
	& Nelson-Aalen  & marginal NA  & Nelson-Aalen  & marginal NA \\
	5 & .949 (.933, .962) & .003 (.001, .009) & .689 (.659, .718) & 0 (0, .004)\\
	10 & .957 (.943, .969) & .004 (.001, .010) & .960 (.946, .971) & 0 (0, .004)\\
	25 & .953 (.938, .965) & .011 (.006, .020) & .963 (.949, .974) & 0 (0, .004)\\
	50 & .951 (.936, .964) & .056 (.043, .072) & .944 (.928, .957) & 0 (0, .004)\\
	75 & .934 (.917, .949) & .842 (.818, .864) & .963 (.949, .970) & .001 (.000, .006)\\
	90 & .932 (.915, .947) & .091 (.074, .111) & .959 (.945, .970) & .178 (.155, .203)\\
	95 & .941 (.925, .955) & .035 (.024, .048) & .945 (.929, .958) & .004 (.001, .010)\\[5pt]
	& Kaplan-Meier  & marginal KM  & Kaplan-Meier  & marginal KM\\
	5 & .950 (.935, .963) & .003 (.001, .009) & .689 (.659, .718) & 0 (0, .004)\\
	10 & .958 (.944, .970) & .004 (.001, .010) & .960 (.946, .971) & 0 (0, .004)\\
	25 & .953 (.938, .965) & .011 (.006, .020) & .963 (.949, .974) & 0 (0, .004)\\
	50 & .951 (.936, .964) & .056 (.043, .072) & .944 (.928, .957) & 0 (0, .004)\\
	75 & .933 (.916, .948) & .842 (.818, .864) & .963 (.949, .974) & .001 (.000, .006)\\
	90 & .930 (.912, .945) & .091 (.074, .111) & .959 (.945, .970) & .178 (.155, .203)\\
	95 & .941 (.925, .955) & .035 (.024, .048) & .938 (.921, .952) & .004 (.001, .010)
    \end{tabular}
    }
\end{table}

\end{document}